\documentclass[11pt]{article}
\def\directunion{\hbox{$\bigcirc$ \hskip - 11.3 pt \raise 0.1pt
\hbox{$\scriptstyle \vee$}}\ }
\newtheorem{theorem}{Theorem}
\newtheorem{definition}{Definition}
\newtheorem{lemma}{Lemma}
\newtheorem{proposition}{Proposition}
\newtheorem{corollary}{Corollary}
\newtheorem{axiom}{Axiom}

\usepackage[psamsfonts]{amssymb}

\newcommand{\qed}{\mbox{} \hfill $\Box$}
\newcommand{\bd}{\begin{definition}}
\newcommand{\ed}{\end{definition}}
\newcommand{\bp}{\begin{proposition}}
\newcommand{\be}{\begin{equation}}
\newcommand{\ee}{\end{equation}}
\newcommand{\bea}{\begin{eqnarray}}
\newcommand{\eea}{\end{eqnarray}}
\newcommand{\ba}{\begin{array}}
\newcommand{\ea}{\end{array}}

\def\zero{\bar\mathtt{0}}
\def\one{\bar\mathtt{1}}
\begin{document}

\title{Decomposition Theorem for State Property Systems}
\author{Diederik Aerts, Didier Deses and Bart D'Hooghe\\
        \normalsize\itshape
        Center Leo Apostel for Interdisciplinary Studies (CLEA) \\
         \normalsize\itshape
         and Foundations of the Exact Sciences (FUND) \\
		\normalsize\itshape
		Department of Mathematics, Vrije Universiteit Brussel \\
        \normalsize\itshape
         1160 Brussels, Belgium \\
        \normalsize
        E-Mails: \textsf{diraerts@vub.ac.be, diddesen@vub.ac.be} \\
		\normalsize
		\textsf{bdhooghe@vub.ac.be}}
\date{}
\maketitle
\begin{abstract}
\noindent We prove a decomposition theorem for orthocomplemented state property systems. More specifically we prove that an orthocomplemented state property system is isomorphic to the direct union of the non classical components of this state property system over the state space of the classical state property system of this state property system.
\end{abstract}

\section{Introduction}
The notion of {\it state property system} \cite{Aerts1999,ACVV1999} was introduced as the mathematical object that represents the set of states and the set of properties of a physical entity following the axiomatic approach to quantum mechanics as developed in Geneva and Brussels \cite{Piron1964,Piron1976,Aerts1981,Aerts1982,Aerts1983a,Piron1990}.
\begin{definition}[state property system]
A state property system is a triple $(\Sigma, {\cal L}, \kappa)$ where $\Sigma$ is a set, ${\cal L}$ is a complete lattice, and $\kappa: {\cal L} \rightarrow {\cal P}(\Sigma)$ is a function, such that
\bea
\kappa(\zero) = \emptyset & & \kappa(\one) = \Sigma \label{eq:cartan01} \\
\kappa(\wedge_ia_i) &=& \cap_i \kappa(a_i)  \label{eq:cartan02}\\
a < b &\Leftrightarrow& \kappa(a) \subset \kappa(b) \label{eq:cartan03}
\eea
\end{definition}
The state property system describes a physical entity, where $\Sigma$ represents the set of states of the physical entity, ${\cal L}$ the set of properties, and $\kappa$ the Cartan map, such that $p \in \kappa(a)$ expresses the following fundamental physical situation: ``The
property $a \in {\cal L}$ is `actual' if the physical entity is in state $p \in \Sigma$".

To capture orthogonality and orthocomplementation in this approach we introduced the notion of ortho state property system, and the notions of ortho couple and ortho property \cite{AertsDeses2005}. Instead of considering the two ortho axioms in \cite{AertsDeses2005} we introduce directly an axiom {\bf OrthoCom} that substitutes for these two axioms.
\begin{axiom} [OrthoCom]
The state property system $(\Sigma, {\cal L}, \kappa)$ satisfies the axiom {\bf OrthoCom} if the lattice ${\cal L}$ of properties of the physical entity under study is
orthocomplemented. This means that there exists a
function $^\perp :{\cal L} \rightarrow {\cal L}$ such that for $a, b \in {\cal
L}$ we have:
\bea
(a^\perp)^\perp &=& a  \label{ortho01}\\
a < b &\Rightarrow& b^\perp < a^\perp \label{ortho02}\\
a \wedge a^\perp = 0\ &{\rm and}&\ a \vee a^\perp = I \label{ortho03}
\eea
For $p, q, \in \Sigma$ we say that $p \perp q$ iff there exists $a \in {\cal L}$ such that $p \in \kappa(a)$ and $q \in \kappa(a^\perp)$. 

Additionally to the orthocomplementation of the lattice of properties we demand that for $a \in {\cal L}$ we have:
\bea 
\{q\ \vert\ q \in \Sigma, q \perp p\ \forall p \in \kappa(a)\} = \kappa(a)^\perp \subset \kappa(a^\perp) \label{orthocartan}
\eea
for axiom {\bf OrthoCom} to be satisfied (remark that $\kappa(a^\perp) \subset \kappa(a)^\perp$ is always satisfied).
\end{axiom}
We define the notions of classical property and classical state as introduced in \cite{AertsDeses2002,AertsDesesVanderVoorde2005}.

\bd [Classical Property and State]
Suppose that $(\Sigma, {\cal L}, \kappa)$ is the state property system
representing a physical
entity, satisfying the axiom {\bf OrthoCom}. We say that a property $a \in {\cal
L}$ is a classical property if
for $p \in \Sigma$ we have
\be
p \in \kappa(a) \ {\rm or}\ p \in \kappa(a^\perp)
\ee
The set of all classical properties we denote by ${\cal C}$.
For $p \in \Sigma$ we introduce
\bea
\omega(p) &=& \bigwedge_{p \in \kappa(a), a \in {\cal C}}a \\
\kappa_c(a) &=& \{\omega(p)\ \vert \ p \in \kappa(a)\}
\eea
and call $\omega(p)$ the classical state of the physical entity 
whenever it is in
a state $p \in \Sigma$, and $\kappa_c$ the classical Cartan map. The set of
all classical states
is denoted by $\Omega$.
\ed

\section{Classical State Property System}
In this section we prove that the set of classical states $\Omega$ and classical properties ${\cal C}$ with the classical Cartan map $\kappa_c$ of a state property system $(\Sigma, {\cal L}, \kappa)$ form a state property system $(\Omega, {\cal C}, \kappa_c)$, which we call the classical state property system of the state property system $(\Sigma, {\cal L}, \kappa)$.

\begin{proposition} \label{classstatepropsystem}
$(\Omega, {\cal C}, \kappa_c)$ is a state property system satisfying the axiom {\bf OrthoCom}. Moreover, each classical state $\omega(p)$ is an atom of the lattice ${\cal C}$, which is an atomistic lattice, and for two classical states $\omega(p), \omega(q) \in \Omega$ we have:
\be
\omega(p) \not= \omega(q) \Rightarrow \omega(p) \perp \omega(q) \label{orthoclassical}
\ee
\end{proposition}
Proof: (i) Consider $a_i \in {\cal C}$, and $p \in \Sigma$ such that $p \not\in \kappa(\wedge_ia_i)=\cap_i\kappa(a_i)$. Then there is $a_j$ such that $p \not\in \kappa(a_j)$. Since $a_j \in {\cal C}$ we have $p \in \kappa(a_j^\perp)$. Hence $p \in \kappa(\vee_ia_i^\perp) = \kappa((\wedge_ia_i)^\perp)$. This proves that $\wedge_ia_i \in {\cal C}$. Take now $q \in \Sigma$ such that $q \not\in \kappa(\vee_ia_i)$. Then $q \not\in \kappa(a_i)\ \forall i$. But as a consequence we have $q \in \kappa(a_i^\perp)\ \forall i$, since $a_i \in {\cal C}\ \forall i$. Hence $q \in \cap_i\kappa(a_i^\perp) = \kappa(\wedge_ia_i^\perp) = \kappa((\vee_ia_i)^\perp)$. This proves that $\vee_ia_i \in {\cal C}$. Obviously $\zero$ and $\one$ are minimal and maximal element of ${\cal C}$. So, we have proven that ${\cal C}$ is a complete lattice.

(ii) $\kappa_c$ is a function from ${\cal L}$ to ${\cal P}(\Omega)$. We have $\kappa_c(\zero) = \emptyset$ and $\kappa_c(\one) = \Omega$, which proves (\ref{eq:cartan01}). Consider $a_i \in {\cal C}$. We have $\kappa_c(\wedge_ia_i) = \{\omega(p)\ \vert\ p \in \kappa(\wedge_ia_i)\} = \{\omega(p)\ \vert\ p \in \cap_i\kappa(a_i)\} = \cap_i\{\omega(p)\ \vert\ p \in \kappa(a_i)\} = \cap_i\kappa_c(a_i)$. This proves (\ref{eq:cartan02}). Consider $a , b \in {\cal L}$ such that $a < b$. Suppose that $\omega(p) \in \kappa_c(a)$, then we have $p \in \kappa(a)$ and hence $p \in \kappa(b)$. As a consequence $\omega(p) \in \kappa_c(b)$. This proves that $\kappa_c(a) \subset \kappa_c(b)$. Suppose now that $a, b \in {\cal C}$ such that $\kappa_c(a) \subset \kappa_c(b)$. Consider $p \in \Sigma$ such that $p \in \kappa(a)$. This means that $\omega(p) \in \kappa_c(a)$ and hence $\omega(p) \in \kappa_c(b)$. As a consequence we have $p \in \kappa(b)$. Hence $\kappa(a) \subset \kappa(b)$ which implies that $a < b$. So we have proven that $\kappa_c$ satisfies (\ref{eq:cartan03}).

(iii) It is easy to check that $\perp$ introduces an orthocomplementation on ${\cal C}$. Before proving (\ref{orthocartan}) we first prove (\ref{orthoclassical}). Consider two classical states $\omega(p)$ and $\omega(q)$ that are different. Since $\omega(p) = \wedge_{p \in \kappa(a), a \in {\cal C}}a$ and $\omega(q) = \wedge_{q \in \kappa(a), a \in {\cal C}}a$, this means that $\{a\ \vert a \in {\cal C}, p \in \kappa(a)\} \not= \{a\ \vert a \in {\cal C}, q \in \kappa(a)\}$. Hence there are two possibilities, (i) there exists an element $a \in {\cal C}$ such that $p \in \kappa(a)$ and $q \not\in \kappa(a)$, or, (ii) there exists an element $b \in {\cal C}$, such that $p \not\in \kappa(b)$ and $q \in \kappa(b)$. This implies that (i) there exist $a \in {\cal C}$ such that $p \in \kappa(a)$ and $q \in \kappa(a^\perp)$, or (ii) there exists $b \in {\cal C}$ such that $p \in \kappa(b^\perp)$ and $q \in \kappa(b)$. As a consequence we have $p \perp q$. Moreover this also implies that (i) $\omega(p) \in \kappa_c(a)$ and $\omega(q) \in \kappa_c(a^\perp)$, or (ii) $\omega(p) \in \kappa_c(b^\perp)$ and $\omega(q) \in \kappa_c(b)$. Hence $\omega(p) \perp \omega (q)$, which proves (\ref{orthoclassical}). 

Consider $\omega(q) \in \kappa_c(a)^\perp$ for $a \in {\cal C}$. This means that for an arbitrary $\omega(p) \in \kappa_c(a)$ we have $\omega(q) \perp \omega(p)$. Hence for an arbitrary $p \in \kappa(a)$ we have $q \perp p$. This means that $q \in \kappa(a)^\perp = \kappa(a^\perp)$. Hence $\omega(q) \in \kappa_c(a^\perp)$. As a consequence we have $\kappa_c(a)^\perp \subset \kappa_c(a^\perp)$, which proves (\ref{orthocartan}).

Let us prove now that each classical state is an atom of the lattice ${\cal C}$. Consider $\omega(p) \in \Omega$, and $a \in {\cal C}$ such that $\zero < a < \omega(p)$ and $\zero \not= a$. Since $\zero \not= a$ there exists $q \in \Sigma$ such that $q \in \kappa(a)$, an hence $\omega(q) < a$. This means that $\omega(q) < \omega(p)$. Suppose that $\omega(q) \not= \omega(p)$, then $\omega(p) \perp \omega(q)$, hence $\omega(q) < \omega(p)^\perp$. Together with $\omega(q) < \omega(p)$ this leads to a contradiction, namely $\omega(q) \in \omega(p) \wedge \omega(p)^\perp = \zero$, which proves that $\omega(p) = \omega(q)$. Hence $a = \omega(p)$, which proves that $\omega(p)$ is an atom of ${\cal C}$. Consider $a \in {\cal C}$ such that $a \not= \zero$. We have $\vee_{\omega(p) < a}\omega(p) < a$. Consider $q \in \Sigma$ such that $q \in \kappa(a)$, then $\omega(q) < a$. This means that $\omega(q) < \vee_{\omega(p) < a}\omega(p)$, and as a consequence $\kappa(\omega(q)) \subset \kappa(\vee_{\omega(p) < a}\omega(p))$. We have $q \in \kappa(\omega(q))$ and hence $q \in \kappa(\vee_{\omega(p) < a}\omega(p))$. This means that we have proven that $\kappa(a) \subset \kappa(\vee_{\omega(p) < a}\omega(p))$, from which follows that $a < \vee_{\omega(p) < a}\omega(p)$. As a consequence $a = \vee_{\omega(p) < a}\omega(p)$, which proves that ${\cal C}$ is an atomistic lattice.
\qed

\begin{definition} [Classical State Property System]
We call $(\Omega, {\cal C}, \kappa_c)$ the classical state property system corresponding with $(\Sigma, {\cal L}, \kappa)$.
\end{definition}
It is possible to prove much more than what we did in proposition \ref{classstatepropsystem}, namely that the classical state property system $(\Omega, {\cal C}, \kappa_c)$ is isomorphic to the canonical state property system $(\Omega, {\cal P}(\Omega), Id)$.

\begin{theorem}
$\kappa_c: {\cal C} \rightarrow {\cal P}(\Omega)$ is an
isomorphism.
\end{theorem}
Proof: From (\ref{eq:cartan03}) follows that $\kappa_c$ is an injective function. Consider $a \in {\cal C}$ and $\kappa_c(a^\perp)$. We know that $\kappa_c(a^\perp) = \kappa_c(a)^\perp$. But, since two classical states are orthogonal from the moment they are different, we have $\kappa_c(a)^\perp = \kappa_c(a)^C$ which is the set theoretical complement of $\kappa_c(a)$ in $\Omega$. Hence $\kappa_c(a^\perp) = \kappa_c(a)^C$. Let us prove that $\kappa_c$ is a surjective function. Take an arbitrary element $A \in {\cal P}(\Omega)$. Consider the property 
\be
a = \bigwedge_{\kappa_c(\omega(p)) \subset A^C} \omega(p)^\perp
\ee
We have
\bea
\kappa_c(a) &=& \kappa_c(\bigwedge_{\kappa_c(\omega(p)) \subset A^C} \omega(p)^\perp) = \bigcap_{\kappa_c(\omega(p)) \subset A^C} \kappa_c(\omega(p)^\perp) = \bigcap_{\kappa_c(\omega(p)) \subset A^C} \kappa_c(\omega(p))^C \\
&=& (\bigcup_{\kappa_c(\omega(p)) \subset A^C} \kappa_c(\omega(p)))^C = (A^C)^C = A
\eea
\qed

\section{Decomposition of a State Property System}
Before we prove the decomposition theorem that we announced we need to prove some specific features of classical properties.

\begin{lemma}
For $x \in {\cal L}$ and $a \in {\cal C}$ we have
\bea
x &=& (x \wedge a) \vee (x \wedge a^\perp) \label{class01} \\
\kappa(x) &=& \kappa(x \wedge a) \cup \kappa(x \wedge a^\perp) \label{class02}
\eea
\end{lemma}
Proof: Since $x \wedge a < x$ and $x \wedge a^\perp < x$ we have $(x \wedge a) \vee (x \wedge a^\perp) < x$. Since $a \in {\cal C}$ we have $\kappa(a^\perp) = \kappa(a)^C$, and hence $\kappa(a) \cup \kappa(a^\perp) = \Sigma$. This gives $\kappa(x) = \kappa(x) \cap (\kappa(a) \cup \kappa(a^\perp)) = (\kappa(x) \cap \kappa(a)) \cup (\kappa(x) \cap \kappa(a^\perp)) = \kappa(x \wedge a) \cup \kappa(x \wedge a^\perp) \subset \kappa((x \wedge a) \vee (x \wedge a^\perp))$. This proves (\ref{class01}) and (\ref{class02}).

\begin{lemma}
For $x, y \in {\cal L}$ and $a \in {\cal C}$ such that $x < a$ and $y < a^\perp$ we have
\bea
(x \vee y)^\perp &=& (x^\perp \wedge a) \vee (y^\perp \wedge a^\perp) \label{class03b} \\
(x \vee y) \wedge a &=& x \label{class03}
\eea
\end{lemma}
Proof: We have $a^\perp < x^\perp$ and $a < y^\perp$. From this follows that $y^\perp \wedge a^\perp < x^\perp$ and $x^\perp \wedge a < y^\perp$. This implies that $x^\perp \wedge y^\perp \wedge a^\perp = y^\perp \wedge a^\perp$ and $x^\perp \wedge y^\perp \wedge a = x^\perp \wedge a$. Since $a \in {\cal C}$ we have $x^\perp \wedge y^\perp = (x^\perp \wedge y^\perp \wedge a) \vee (x^\perp \wedge y^\perp \wedge a^\perp)$. So $x^\perp \wedge y^\perp = (x^\perp \wedge a) \vee (y^\perp \wedge a^\perp)$. Hence $x \vee y = (x \vee a^\perp) \wedge (y \vee a)$. But then $(x \vee y) \wedge a = (x \vee a^\perp) \wedge a$. We know that $x^\perp = (x^\perp \wedge a) \vee (x^\perp \wedge a^\perp) = (x^\perp \wedge a) \vee a^\perp$. Hence $x = (x \vee a^\perp) \wedge a$. This proves that $(x \vee y) \wedge a = x$.
\qed

\begin{lemma}
For $x, x_i \in {\cal L}$ and $a \in {\cal C}$ we have
\bea
&& a \wedge (\vee_ix_i) = \vee_i(a \wedge x_i) \label{class04} \\
&& a = (a \wedge x) \vee (a \wedge x^\perp) \label{class04b}
\eea
\end{lemma}
Proof: We have $a \wedge (\vee_ix_i) = a \wedge (\vee_i((x_i \wedge a) \vee (x_i \wedge a^\perp)) = a \wedge (\vee_i(x_i \wedge a) \vee \vee_i(x_i \wedge a^\perp)) = \vee_i(x_i \wedge a)$. We have $a = a \wedge (x \vee x^\perp)$. From (\ref{class04}) follows that $a \wedge (x \vee x^\perp) = (a \wedge x) \vee (a \wedge x^\perp)$, which proves (\ref{class04b}).
\qed

\begin{lemma}
For $a \in {\cal L}$ we have
\bea
a &=& \bigvee_{\omega \in \Omega} (a \wedge \omega) \label{class05} \\
\kappa(a) &=& \bigcup_{\omega \in \Omega}\kappa(a \wedge \omega) \label{class06}
\eea
with
\be
a \wedge \omega \perp a \wedge \omega'\quad {\rm and}\quad \kappa(a \wedge \omega) \cap \kappa(a \wedge \omega') = \emptyset\quad {\rm for}\quad \omega \not= \omega'
\ee
\end{lemma}
Proof:  We have that $a \wedge \omega < a\ \forall \omega \in \Omega$, hence $\kappa(a \wedge \omega) \subset \kappa(a)\ \forall\ \omega \in \Omega$, and as a consequence $\cup_{\omega \in \Omega}\kappa(a \wedge \omega) \subset \kappa(a)$. Consider $p \in \kappa(a)$. We have $p \in \kappa(\omega(p))$, and hence $p \in \kappa(a) \cap \kappa(\omega(p)) = \kappa(a \wedge \omega(p)) \subset \cup_{\omega \in \Omega}\kappa(a \wedge \omega)$. So we have shown that $\kappa(a) \subset \cup_{\omega \in \Omega}\kappa(a \wedge \omega)$. This proves (\ref{class06}), namely $\kappa(a) = \cup_{\omega \in \Omega}\kappa(a \wedge \omega)$. We have that $a \wedge \omega < a\ \forall \omega \in \Omega$, hence $\vee_{\omega \in \Omega} (a \wedge \omega) < a$. Consider $p \in \kappa(a)$. We have $p \in \cup_{\omega \in \Omega}\kappa(a \wedge \omega) \subset \kappa(\vee_{\omega \in \Omega}(a \wedge \omega))$. So we have shown that $\kappa(a) \subset \kappa(\vee_{\omega \in \Omega}(a \wedge \omega))$. From this follows that $a < \vee_{\omega \in \Omega}(a \wedge \omega)$, which proves (\ref{class05}), namely $a = \vee_{\omega \in \Omega}(a \wedge \omega)$. Consider $\omega \not= \omega'$, then we have $\omega < \omega'^\perp$. As a consequence $a \wedge \omega < \omega'^\perp < a^\perp \vee \omega'^\perp = (a \wedge \omega')^\perp$, which proves that $a \wedge \omega \perp a \wedge \omega'$. From this follows that $\kappa(a \wedge \omega) \cap \kappa(a \wedge \omega') = \emptyset$.

\qed

\begin{corollary}
We have 
\be
\Sigma = \bigcup_{\omega \in \Omega}\kappa(\omega) \label{class07}
\ee
with
\be
\kappa(\omega) \cap \kappa(\omega') = \emptyset\quad {\rm for}\quad \omega \not= \omega'
\ee
\end{corollary}

\begin{lemma}
Consider $a_\omega$ such that $a_\omega < \omega\ \forall\ \omega \in \Omega$. We have
\be
\kappa(\bigvee_{\omega \in \Omega}a_\omega) = \bigcup_{\omega \in \Omega}\kappa(a_\omega) \label{class08}
\ee
with
\be
\kappa(a_\omega) \cap \kappa(a_{\omega'}) = \emptyset\quad {\rm for}\quad \omega \not= \omega'
\ee
\end{lemma}
Proof: We have $\kappa(\vee_{\omega \in \Omega}a_\omega) = \cup_{\omega' \in \Omega}\kappa((\vee_{\omega \in \Omega}a_\omega) \wedge \omega')$. From (\ref{class03}) follows that $(\vee_{\omega \in \Omega}a_\omega) \wedge \omega' =  a_{\omega'}$. Hence $\kappa(\vee_{\omega \in \Omega}a_\omega) = \cup_{\omega' \in \Omega}\kappa(a_{\omega'})$. This proves (\ref{class08}). 
\qed

\medskip
\noindent
Let us now investigate the nonclassical parts of the state property system $(\Sigma, {\cal L}, \kappa)$.

\bd [Nonclassical Part] \label{def:nonclassicalcomponents}
Suppose that $(\Sigma, {\cal L}, \kappa)$ is the state property system of a
physical entity satisfying the axiom {\bf OrthoCom}. For $\omega \in \Omega$ we introduce
\bea
{\cal L}_\omega &=& \{a\ \vert a < \omega, a \in {\cal L}\} \\
\Sigma_\omega &=& \{p\ \vert p \in \kappa(\omega), p \in \Sigma\} \\
\kappa_\omega(a) &=& \kappa(a)\ {\rm for}\ a \in {\cal L}_\omega
\eea
and for $a \in {\cal L}_\omega$ we define
\be
a^{\perp_\omega} = a^\perp \wedge \omega \label{omegaortho}
\ee
and we call $(\Sigma_\omega, {\cal L}_\omega, \kappa_\omega)$ the nonclassical
components of $(\Sigma, {\cal L}, \kappa)$ corresponding to $\omega$.
\ed

\begin{proposition}
For $\omega \in \Omega$ we have that $(\Sigma_\omega, {\cal L}_\omega, \kappa_\omega)$ is a state property system that satisfies the axiom {\bf OrthoCom}.
\end{proposition}
Proof: It is easy to prove that ${\cal L}_\omega$ is a complete lattice with minimal element $\zero$ and maximal element $\omega$. For $a_i \in {\cal L}_\omega$ we have $\kappa_\omega(\zero) = \emptyset$ and $\kappa_\omega(\omega) = \Sigma_\omega$, and $\kappa_\omega(\wedge_ia_i) = \kappa(\wedge_ia_i) = \cap_i\kappa(a_i) = \cap_i\kappa_\omega(a_i)$, which proves that $\kappa_\omega$ satisfies (\ref{eq:cartan01}) and (\ref{eq:cartan02}). Consider $a , b \in {\cal L}_\omega$, then we have $a < b \Leftrightarrow \kappa(a) \subset \kappa(b) \Leftrightarrow \kappa_\omega(a) \subset \kappa_\omega(b)$, which proves (\ref{eq:cartan03}).

Let us prove that (\ref{omegaortho}) defines an orthocomplementation for which the axiom {\bf OrthoCom} is satisfied for ${\cal L}_\omega$. We have $(a^{\perp_\omega})^{\perp_\omega} = (a^\perp \wedge \omega)^{\perp_\omega} = (a^\perp \wedge \omega)^\perp \wedge \omega = ((a^\perp)^\perp \vee \omega^\perp) \wedge \omega = (a \vee \omega^\perp) \wedge \omega = a$. This proves (\ref{ortho01}). Consider $a, b \in {\cal L}_\omega$ such that $a < b$. We have $b^{\perp_\omega} = b^\perp \wedge \omega < a^\perp \wedge \omega = a^{\perp_\omega}$. This proves (\ref{ortho02}). For $a \in {\cal L}_\omega$ we have $a \wedge a^{\perp_\omega} = a \wedge a^\perp \wedge \omega = \zero$. Further $a \vee a^{\perp_\omega} = a \vee (a^\perp \wedge \omega)$. Following (\ref{class04}) we have $a \vee (a^\perp \wedge \omega) = (a \vee a^\perp) \wedge \omega = \omega$, which proves (\ref{ortho03}).

For $p , q \in \Sigma_\omega$ we have $p \perp_\omega q$ iff there exists $a \in {\cal L}_\omega$ such that $p \in \kappa_\omega(a)$ and $q \in \kappa_\omega(a^{\perp_\omega})$. This means that $p \in \kappa(a)$ and $q \in \kappa(a^\perp) \wedge \omega$, and hence $p \perp q$. Suppose that $p, q \in \Sigma_\omega$ such that $p \perp q$. This means that there exists $a \in {\cal L}$ such that $p \in \kappa(a)$ and $q \in \kappa(a^\perp)$. Since $p, q \in \kappa(\omega)$ we have that $p \in \kappa(a \wedge \omega) = \kappa_\omega(a \wedge \omega)$ and $q \in \kappa(a^\perp \wedge \omega) = \kappa_\omega(a^\perp \wedge \omega)$. We have $(a \wedge \omega)^{\perp_\omega} = (a \wedge \omega)^\perp \wedge \omega = (a^\perp \vee \omega^\perp) \wedge \omega =   (a^\perp \wedge \omega)$. This shows that $p \perp_\omega q$. We have proven that for $p, q \in \Sigma_\omega$ we have $p \perp_\omega q \Leftrightarrow p \perp q$. For $a \in {\cal L}_\omega$ we have $\kappa_\omega(a)^{\perp_\omega} = \kappa(a)^\perp \cap \Sigma_\omega \subset \kappa(a^\perp) \cap \kappa(\omega) = \kappa(a^\perp \wedge \omega) = \kappa(a^{\perp_\omega}) = \kappa_\omega(a^{\perp_\omega})$. This proves (\ref{orthocartan}).
\qed

\medskip
\noindent
To see in
more detail in which way the classical and nonclassical parts are
structured within the lattice ${\cal L}$, we need to introduce some additional structures.

\bd [Direct Union of State Property Systems] Consider a set of state property systems $(\Sigma_\omega, {\cal
L}_\omega, \kappa_\omega)$ that all satisfy the axiom {\bf OrthoCom}. The direct union
$\directunion_\omega (\Sigma_\omega, {\cal L}_\omega, \kappa_\omega)$ of
these state property systems is the
state property system $(\cup_\omega\Sigma_\omega, \directunion_\omega{\cal
L}_\omega,
\directunion_\omega\kappa_\omega)$, where 

\noindent
(i) $\cup_\omega\Sigma_\omega$ is the
disjoint union of the sets
$\Sigma_\omega$

\noindent
(ii) $\directunion_\omega{\cal L}_\omega$ is the direct union
of the lattices ${\cal L}_\omega$, which means the set of sequences $a = (a_\omega)_\omega$, such
that\bea(a_\omega)_\omega < (b
_\omega)_\omega &\Leftrightarrow& a_\omega < b_\omega\ \forall \omega \in
\Omega \\
(a_\omega)_\omega \wedge (b_\omega)_\omega &=& (a_\omega \wedge
b_\omega)_\omega  \\
(a_\omega)_\omega \vee (b_\omega)_\omega &=& (a_\omega \vee b_\omega)_\omega \\
(a_\omega)_\omega^\perp &=& (a_\omega^{\perp_\omega})_\omega
\eea
\noindent
(iii) $\directunion_\omega\kappa_\omega$ is defined as follows:
\be
\directunion_\omega\kappa_\omega ((a_{\omega'})_{\omega'}) =
\cup_\omega\kappa_\omega(a_\omega)
\ee
\ed
Remark that if ${\cal L}_\omega$ are complete orthocomplemented lattices, then also $\directunion_{\omega
\in \Omega}{\cal L}_\omega$ is a complete orthocomplemented lattice. A fundamental decomposition theorem can now be proven.

\begin{theorem} [Decomposition Theorem] 
Consider the state property system $(\Sigma, {\cal L}, \kappa)$, and suppose
that the axiom {\bf OrthoCom} is satisfied. Then
\be
(\Sigma, {\cal L}, \kappa) \cong \directunion_{\omega \in
\Omega}(\Sigma_\omega, {\cal L}_\omega,
\kappa_\omega)
\ee
where $\Omega$ is the set of classical states of $(\Sigma, {\cal L},
\kappa)$, $\Sigma_\omega$ is the set
of states, $\kappa_\omega$ the corresponding Cartan map, and
${\cal L}_\omega$ the lattice of properties of the
nonclassical component
$(\Sigma_\omega, {\cal L}_\omega,
\kappa_\omega)$.
\end{theorem}
Proof: We use the notion of morphism of state property systems as introduced in \cite{ACVV1999}, and need to prove that there exists an ismomorphism of state property systems between $(\Sigma, {\cal L}, \kappa)$ and $\directunion_{\omega \in
\Omega}(\Sigma_\omega, {\cal L}_\omega,
\kappa_\omega)
$. More concretely, consider two state property systems $(\Sigma, {\cal L}, \kappa)$ and $(\Sigma', {\cal L}', \kappa')$. We say that $(m, n): (\Sigma', {\cal L}', \kappa') \rightarrow (\Sigma, {\cal L}, \kappa)$ is a morphism, if $m$ is a function:
\be
m: \Sigma' \rightarrow \Sigma
\ee
and $n$ is a function
\be
n: {\cal L} \rightarrow {\cal L}'
\ee
such that for $a \in {\cal L}$ and $p' \in \Sigma'$ the following holds:
\be
m(p') \in \kappa(a) \Leftrightarrow p' \in \kappa'(n(a)) \label{morphism}
\ee
We say that $(m, n)$ is an isomorphism if $(m, n)$ is a morphism and $m$ and $n$ are bijective. The set of states of the state property system $\directunion_{\omega \in
\Omega}(\Sigma_\omega, {\cal L}_\omega,
\kappa_\omega)$ is given by the disjoint union $\cup_{\omega \in \Omega}\Sigma_\omega$ of sets of states $\Sigma_\omega$ of the nonclassical component state property systems $(\Sigma_\omega, {\cal L}_\omega, \kappa_\omega)$ of the state property system $(\Sigma, {\cal L}, \kappa)$. From (\ref{class07}) follows that $m$ can be defined in the following way:
\bea
m: \Sigma &\rightarrow& \cup_{\omega \in \Omega}\Sigma_\omega \\
p &\mapsto& p 
\eea
The function $n$ is defined in the following way:
\bea
n: \directunion_{\omega \in \Omega}{\cal L}_\omega &\rightarrow& {\cal L}  \\
(a_\omega)_\omega &\mapsto& \vee_{\omega \in \Omega}a_\omega
\eea
The function $m$ is a bijection by definition. Consider $(a_\omega)_\omega, (b_\omega)_\omega \in \directunion_{\omega \in \Omega}{\cal L}_\omega$ and suppose that $n((a_\omega)_\omega) = n((b_\omega)_\omega)$, hence $\vee_{\omega \in \Omega}a_\omega = \vee_{\omega \in \Omega}b_\omega$. Then $(\vee_{\omega \in \Omega} a_\omega) \wedge \omega' = (\vee_{\omega \in \Omega}b_\omega) \wedge \omega'\ \forall\ \omega' \in \Omega$. From (\ref{class03}) follows that $(\vee_{\omega \in \Omega}a_\omega) \wedge \omega' = a_{\omega'}$ and $(\vee_{\omega \in \Omega}b_\omega) \wedge \omega' = b_{\omega'}$. Hence $a_{\omega'} = b_{\omega'}\ \forall\ \omega' \in \Omega$. As a consequence we have $(a_\omega)_\omega = (b_\omega)_\omega$. This proves that $n$ is injective. Let us prove that $n$ is surjective. Consider an arbitrary element $a \in {\cal L}$. From (\ref{class05}) it follows that $a = \vee_{\omega \in \Omega}(a \wedge \omega)$. Consider the element $(a \wedge \omega)_\omega \in \directunion_{\omega \in \Omega}{\cal L}_\omega$. Then $n((a \wedge \omega)_\omega) = a$ which proves that $n$ is surjective. 

Hence we have proven that $m$ as well as $n$ are bijections. Let us show that we have a morphism. We need to prove (\ref{morphism}), hence: 
\be
m(p) \in \directunion_\omega\kappa_\omega((a_{\omega'})_{\omega'}) \Leftrightarrow p \in \kappa(n((a_{\omega'})_{\omega'})) \label{morphismdecomposition}
\ee
We have $m(p) = p$. Let us calculate $\directunion_\omega\kappa_\omega((a_{\omega'})_{\omega'}) = \cup_{\omega \in \Omega}\kappa_\omega(a_\omega) = \cup_{\omega \in \Omega}\kappa(a_\omega)$. On the other hand we have $\kappa(n((a_{\omega'})_{\omega'})) = \kappa(\vee_{\omega \in \Omega}a_\omega)$, and following (\ref{class08}), we have $\kappa(\vee_{\omega \in \Omega}a_\omega) = \cup_{\omega \in \Omega}\kappa(a_\omega)$. This means that (\ref{morphismdecomposition}) is satisfied, and hence $(m, n)$ is an isomorphism of state property systems.

Let us prove additionally that $(m, n)$ also preserves the orthocomplementation structure of $(\Sigma, {\cal L}, \kappa)$ such that it is not only an isomorphism of state property systems, but also an isomorphism for the orthocomplementation structure. We have $n((a_\omega)_\omega^\perp) = n((a_\omega^{\perp_\omega})_\omega) = \vee_{\omega \in \Omega}a_\omega^{\perp_\omega} = \vee_{\omega \in \Omega}(a_\omega^\perp\wedge\omega)$. We also have $n((a_\omega)_\omega)^\perp = (\vee_{\omega \in \Omega}a_\omega)^\perp$. To prove that $n((a_\omega)_\omega^\perp) = n((a_\omega)_\omega)^\perp$ and hence $(m, n)$ is a morphism that also conserves the orthocomplementation, we need to prove that $(\vee_{\omega \in \Omega}a_\omega)^\perp = \vee_{\omega \in \Omega}(a_\omega^\perp\wedge\omega)$, which follows from (\ref{class03b}).
\qed

\section{Conclusion}
We have shown how an orthocomplemented state property system can be
decomposed into a direct union of its non classical components over the
state space of the classical state property system defined by the classical properties of this state property system.
The decomposition theorem tells us the specific way in which classical properties and states can be separated from the non classical components of the state property system of a physical entity and the prominent role played by the orthocomplementation on the lattice of properties.

\end{document}